# Deep prior-based denoising for state-of-the-art scientific imaging and metrology

Yuichi Yokoyama[1*], Kohei Yamagami[1], Yuta Sumiya[2], Hayaru Shouno[2], and Masaichiro Mizumaki[3*]

[1]*Japan Synchrotron Radiation Research Institute (JASRI), Sayo, Hyogo 679-5198, Japan.*
[2]*The University of Electro-Communications, Department of Informatics, Chofu, Tokyo 182-8585, Japan.*
[3]*Faculty of Science, Kumamoto University, Kurokami, Kumamoto 860-8555, Japan.*
*Corresponding authors:
Yuichi Yokoyama (y.yokoyama@spring8.or.jp), Masaichiro Mizumaki (mizumaki@kumamoto-u.ac.jp)

Deep learning has revolutionized computer vision, yet a major gap persists between complex, data-hungry deep learning models and the practical demands of state-of-the-art scientific measurements. To fundamentally bridge this gap, we propose deep prior-based denoising, a robust deep learning model that requires no training data. We demonstrate its effectiveness by removing grid artifacts in angle-resolved photoemission spectroscopy (ARPES), a long-standing and critical data analysis challenge in materials science. Our results demonstrate that deep prior-based denoising yields clearer ARPES images in a fraction of the time required by conventional, experiment-based denoising methods. This ultra-efficient approach to ARPES will enable high-speed, high-resolution three-dimensional band structure mapping in momentum space, thereby dramatically accelerating our understanding of microscopic electronic structures of materials. Beyond ARPES, deep prior-based denoising represents a versatile tool that could become a new standard in any advanced scientific measurement fields where data acquisition is limited.

**Introduction**

The advent of deep learning, particularly convolutional neural networks (CNNs), has profoundly advanced image recognition[1]. Subsequent architectures, typified by U-Net[2], have enabled precise image segmentation even with limited datasets, yielding significant benefits in fields such as medical diagnostics. However, state-of-the-art performance is achieved by large-scale deep learning models such as Vision Transformer (ViT)[3], which are inspired by natural language processing and trained on image datasets comprising hundreds of millions of samples.

This reliance on data-intensive deep learning models has created a mismatch with the practical demands at the forefront of scientific discovery. For example, in advanced measurements using quantum beams such as synchrotron radiation X-rays or neutrons, the immense cost per experiment makes it impractical to build the large-scale datasets required for training deep learning models. Furthermore, a significant hurdle exists for experts in advanced instrumentation to optimize and integrate these complex deep learning models into their experimental systems. Consequently, many advanced scientific measurement fields have yet to fully utilize the revolutionary image recognition capabilities of deep learning. To fundamentally overcome these challenges, we propose deep prior-based denoising, an image reconstruction method that utilizes a simple deep learning model yet requires no training data. This method leverages the inherent deep prior embedded in the structure of CNN, enabling a high-precision restoration of signal-only image from a single image containing noise and artifacts. This approach capitalizes on the inductive bias of CNNs, a feature lacking in ViT, to eliminate the need for large datasets, offering excellent practicality in terms of computation time and ease of implementation. Deep prior-based denoising holds strong potential to become a new standard in any scientific measurement fields where data acquisition is limited.

In this paper, we demonstrate the effectiveness of deep prior-based denoising through a case study on angle-resolved photoemission spectroscopy (ARPES), addressing a critical challenge in materials science. ARPES is a powerful experimental technique that probes the electronic states of materials in reciprocal space and has been instrumental in studying high-temperature superconductors, topological insulators, and two-dimensional materials[4–6]. To understand the band structure, which is described as a function of the kinetic energy and momentum of electrons, it is essential to map it across the three-dimensional momentum space. This mapping requires the measurement of numerous two-dimensional ARPES images. However, even under ultra-high vacuum conditions, the entire series of measurements must be completed rapidly to prevent sample degradation over time because ARPES is a surface-sensitive technique. Furthermore, while high resolution is necessary to obtain detailed information about the band structure, the resolution and the signal-to-noise ratio (S/N) are in a trade-off relationship, with high-resolution measurements leading to a lower S/N. The use of soft X-rays, which have higher energy than conventional VUV light, enables the observation of a wide momentum space covering multiple Brillouin zones[7], but soft X-rays further degrade the S/N because their photoionization cross-section is two orders of magnitude smaller[8]. Moreover, the S/N has become a primary bottleneck in the latest frontiers of ARPES, such as spin-resolved ARPES, time-resolved ARPES, and momentum microscopy[6], creating a strong demand for image-processing technologies to enhance signal quality.

In conventional high-resolution ARPES measurements, the swept mode of a hemispherical electron energy analyzer is commonly used. This method acquires data over a wide energy range by sweeping the detector voltage. However, the swept mode is highly inefficient. To average the signal across the detector for a specific energy range of interest, the voltage sweep must extend into extraneous energy regions, wasting valuable measurement time. Alternatively, a more time-efficient fixed mode is available for measurements within an approximately 5–10 eV range, which offers a significantly higher S/N than the swept mode. Nevertheless, the critical drawback of the fixed mode is the emergence of grid-like artifacts, which are fundamentally different from statistical random noise. These grid artifacts originate from a metal mesh installed in the photoelectron analyzer. The grid artifacts throughout an ARPES image obscure the underlying band structures and compromises ARPES spectra, such as energy distribution curves (EDCs) and momentum distribution curves (MDCs), often rendering them unusable for analysis.

To enhance the efficiency of ARPES measurements, two distinct strategies exist. The first strategy is to enhance the S/N of the swept mode data by removing statistical random noise. Various denoising techniques have been investigated for this purpose, including classical Gaussian filtering, total variation (TV) regularization denoising[9], or deep learning methods[10–13]. The second, more ambitious strategy is to process the high-S/N fixed mode data, addressing both random noise and the systematic grid artifacts simultaneously. We adopt this latter strategy, capitalizing on the superior intrinsic signal quality of the fixed mode. Removing the grid artifacts is a significantly more complex problem than simple denoising. While early efforts used Fourier transform methods[14], a recent deep learning approach achieve higher-precision results[15] but face practical limitations: This deep learning method, however, suffers from practical drawbacks. It requires a complex combination of multiple deep learning models, making control difficult due to the need to balance learning rates delicately. The associated computational overhead also makes it ill-suited for real-time integration into experiments. Our deep prior-based denoising method overcomes these limitations. It simultaneously eliminates the grid artifacts and random noise using a single, unified deep learning model, which radically simplifies hyperparameter management while operating at high speed. Here, we demonstrate that this streamlined and powerful approach enables a new regime of ultra-efficient ARPES measurements.

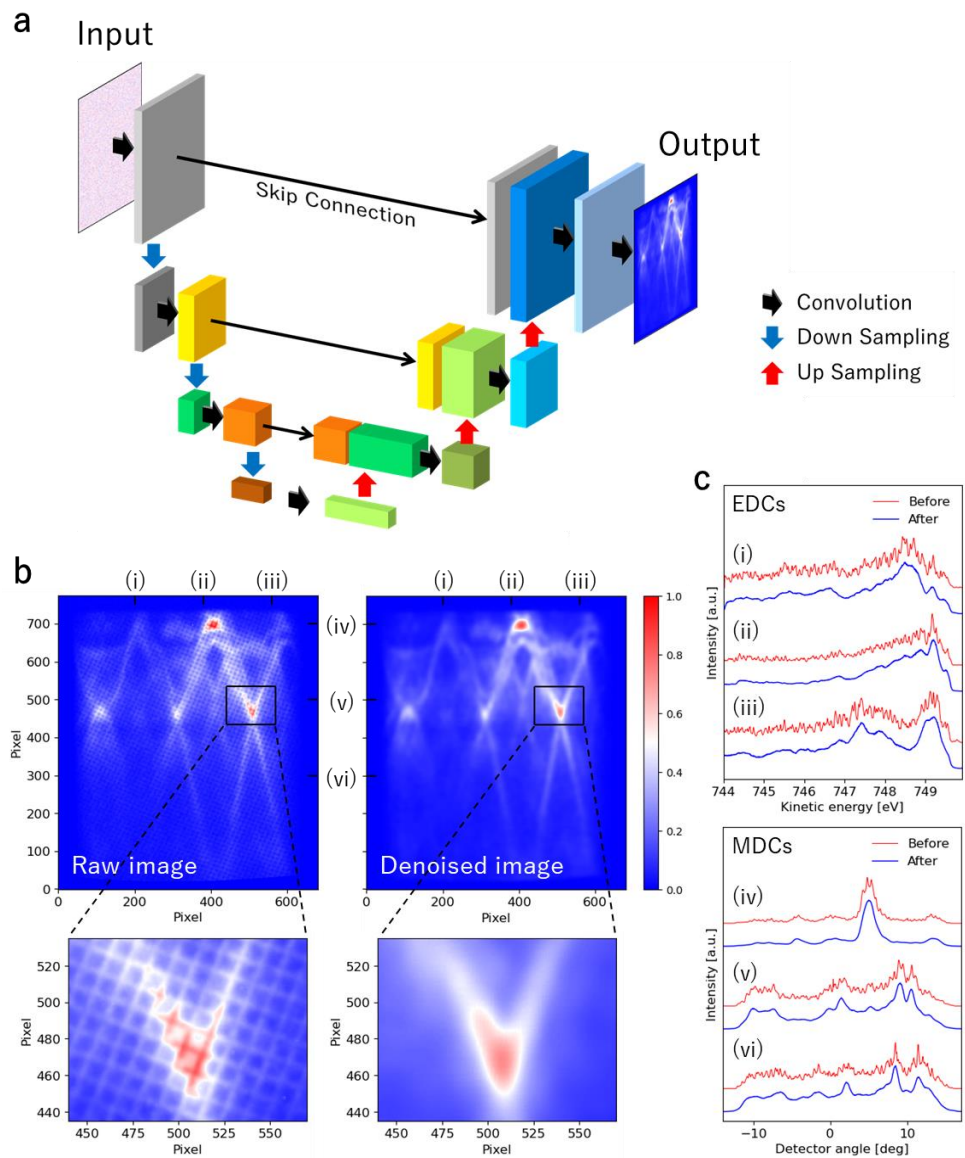

**Fig. 1: Deep prior-based denoising for soft X-ray ARPES. a**, Network architecture for deep prior-based denoising. **b**, Raw image acquired in fixed mode ARPES compared to the denoised image. Expanded views of the rectangular regions are also shown. **c**, Energy distribution curves (EDCs) and momentum distribution curves (MDCs) before and after deep prior-based denoising.

## Results

### Deep prior and its application to soft X-ray ARPES

We developed a deep learning model, termed "Deep prior", to separate the ARPES signal, the grid artifacts, and noise from the raw ARPES image. Our design was motivated by prior findings that CNNs learn structured components more rapidly than unstructured noise[13]. We extended this principle to the more complex ARPES problem, hypothesizing that not only noise but also the ARPES signal and the grid artifacts have distinct learning speeds within a neural network.

The resulting architecture is a four-layer U-shaped CNN with skip connections (U-net), shown in Fig. 1a. Deep prior leverages the neural network architecture itself as an implicit prior, reconstructing the experimentally acquired raw ARPES image without requiring any training data. The input to the U-Net is a random image sampled from a uniform distribution, $U[0,1]$, and the network parameters are learned starting from random values. The Deep prior is learned from scratch on each ARPES image, ensuring high generalizability and adaptability to novel data.

We used the mean squared error (MSE) between the U-Net output and the target ARPES image as the loss function. The network will eventually learn to perfectly reproduce the target image, including the grid artifacts and noise. However, this overfitting is addressed by implementing an early stopping strategy that leverages the plateau of the loss function, as will be detailed later. As a proof of principle, we demonstrated the efficacy of deep prior-based denoising on experimental data from soft X-ray ARPES (SX-ARPES) measurements of $CeRu_2Si_2$, a canonical heavy-fermion compound. SX-ARPES data is well-suited for evaluating the performance of deep prior-based denoising, as its photoionization cross-section is two orders of magnitude lower than that of VUV-ARPES, resulting in poor statistical precision. Figure 1b compares the raw SX-ARPES image, acquired in just 40 s, with the denoised result. In the raw image, the pervasive grid artifacts obscure the band structure. After denoising, the grid artifacts are effectively removed, revealing the band structure clearly. This improvement is also visible in the expanded view of the rectangular region.

Figure 1c presents the EDCs and MDCs before and after denoising. Before denoising (red lines), intensity fluctuations caused by the grid artifacts and noise obscure the intrinsic spectral shapes. After denoising (blue lines), the spectra are smooth and free of the grid artifacts and noise, allowing accurate observation of the true EDC and MDC profiles.

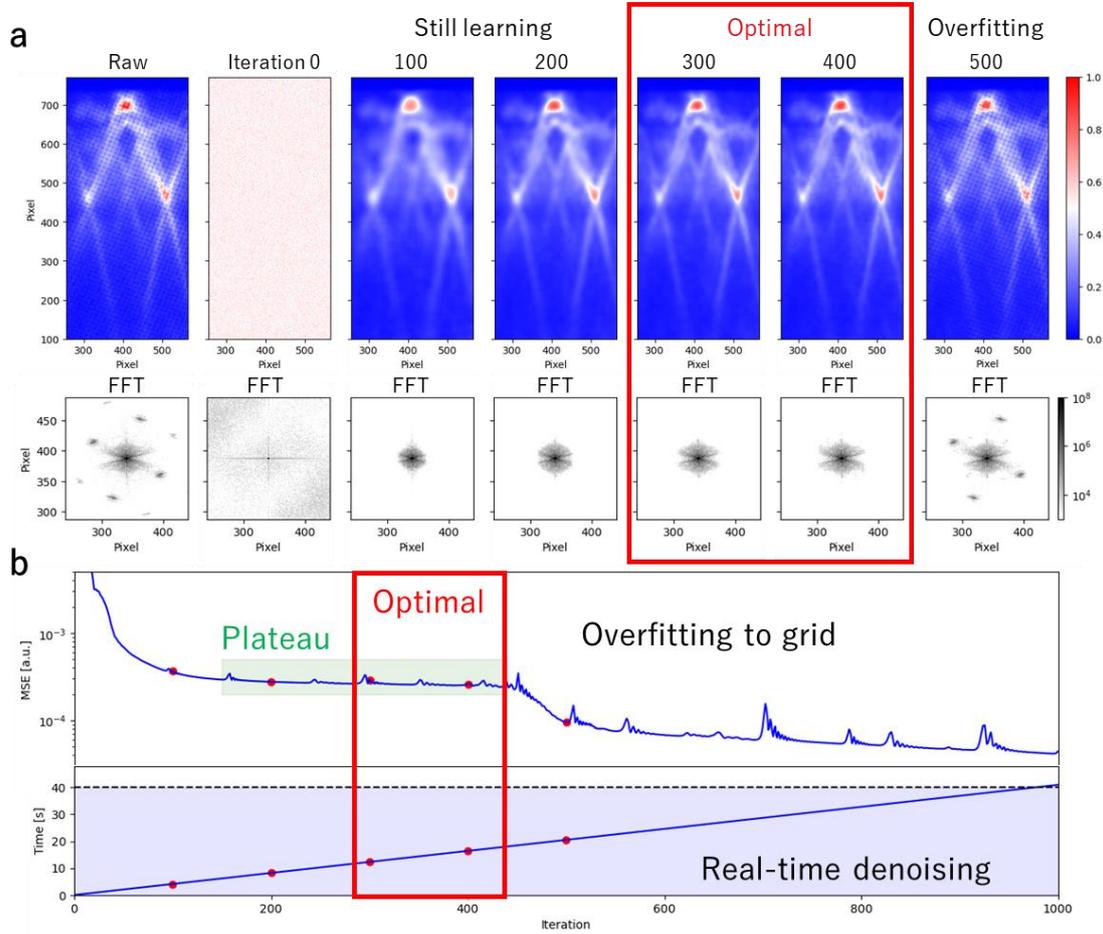

**Fig. 2: Operating principle of deep prior-based denoising. a**, Raw ARPES image, output images, and corresponding fast Fourier transform (FFT) images at 0, 100, 200, 300, 400, and 500 iterations. **b**, MSE learning curve and denoising time. Real-time denoising refers to processing performed during ARPES measurements.

**Operating principle of deep prior-based denoising for ultra-high-throughput ARPES**

We describe the operating principle of deep prior-based denoising, which extracts the intrinsic ARPES signal by separating it from the grid artifacts and noise. The method employs an MSE loss function, defined as:

$$loss = MSE\{f_{\boldsymbol{\theta}}(\boldsymbol{z}), \boldsymbol{x}_{ARPES}\}, \qquad (1)$$

where $\boldsymbol{x}_{ARPES}$ is the target ARPES image, and $f_{\boldsymbol{\theta}}(\boldsymbol{z})$ is the output of the U-net with parameters $\boldsymbol{\theta}$ from the input image $\boldsymbol{z}$. Direct optimization of this loss function would reproduce the target image along with the grid artifacts and noise, making signal isolation challenging. However, due to the spectral bias of CNNs[16], low-frequency structures are learned faster than high-frequency ones. Consequently, the learning proceeds sequentially: first the signal, then the grid artifacts, and finally noise. This property enables effective early stopping to recover a clean ARPES signal free of artifacts and noise.

Figure 2 illustrates our early stopping strategy. Output images at 0, 100, 200, 300, 400, and 500 iterations, along with their frequency-space representations via fast Fourier transform (FFT), are shown in Fig. 2a. Initially, the output image consists primarily of random noise. By 100 iterations, the band structure begins to emerge, but it remains blurry, indicating that the APRES signal has not yet been fully learned. Between 200 and 400 iterations, a band structure becomes sharply defined and free of the grid artifacts. Corresponding FFT images reveal that learning progresses from low to higher frequencies, stabilizing between 300 and 400 iterations, which defines the optimal early stopping. At 500 iterations, the grid artifacts reappear, mirrored by four characteristics of FFT spots matching those in the raw ARPES image, indicating overfitting.

The MSE learning curve in Fig. 2b further illustrates this progression. The curve exhibits an initial steep decline followed by a prolonged plateau from approximately 150 to 440 iterations. Beyond this plateau, the loss decreases again as the network begins to overfit the grid artifacts. Since the grid artifacts are absent before 400 iterations, the plateau region corresponds to signal reconstruction that excludes the grid artifacts, thereby defining the optimal early stopping window. Halting the learning within the region highlighted by the red box effectively isolates the ARPES signal from the grid artifacts and noise.

Denoising on an NVIDIA RTX A6000 GPU required less than 20 s to reach the end of the plateau. Given that a typical ARPES measurement takes several minutes to an hour, this method completes denoising in a fraction of the acquisition time. Even for challenging case, such as an image acquired in 40 s, denoising completes in under half the measurement duration.

Ultimately, ARPES aims to map band structures in three-dimensional momentum space, requiring the continuous acquisition of numerous ARPES images while varying experimental conditions such as incident photon energy and sample orientation. With the conventional swept mode, this goal has been largely impractical, as the long measurement time required for adequate statistics and energy resolution is incompatible with the need to prevent sample surface degradation. Our deep prior-based denoising, when combined with the efficient fixed mode, fundamentally changes this paradigm. It enables an ultra-high-throughput ARPES methodology by drastically reducing the time required for each ARPES image. This work thus illuminates a direct pathway to realizing the long-standing goal of rapid, high-resolution 3D band structure mapping.

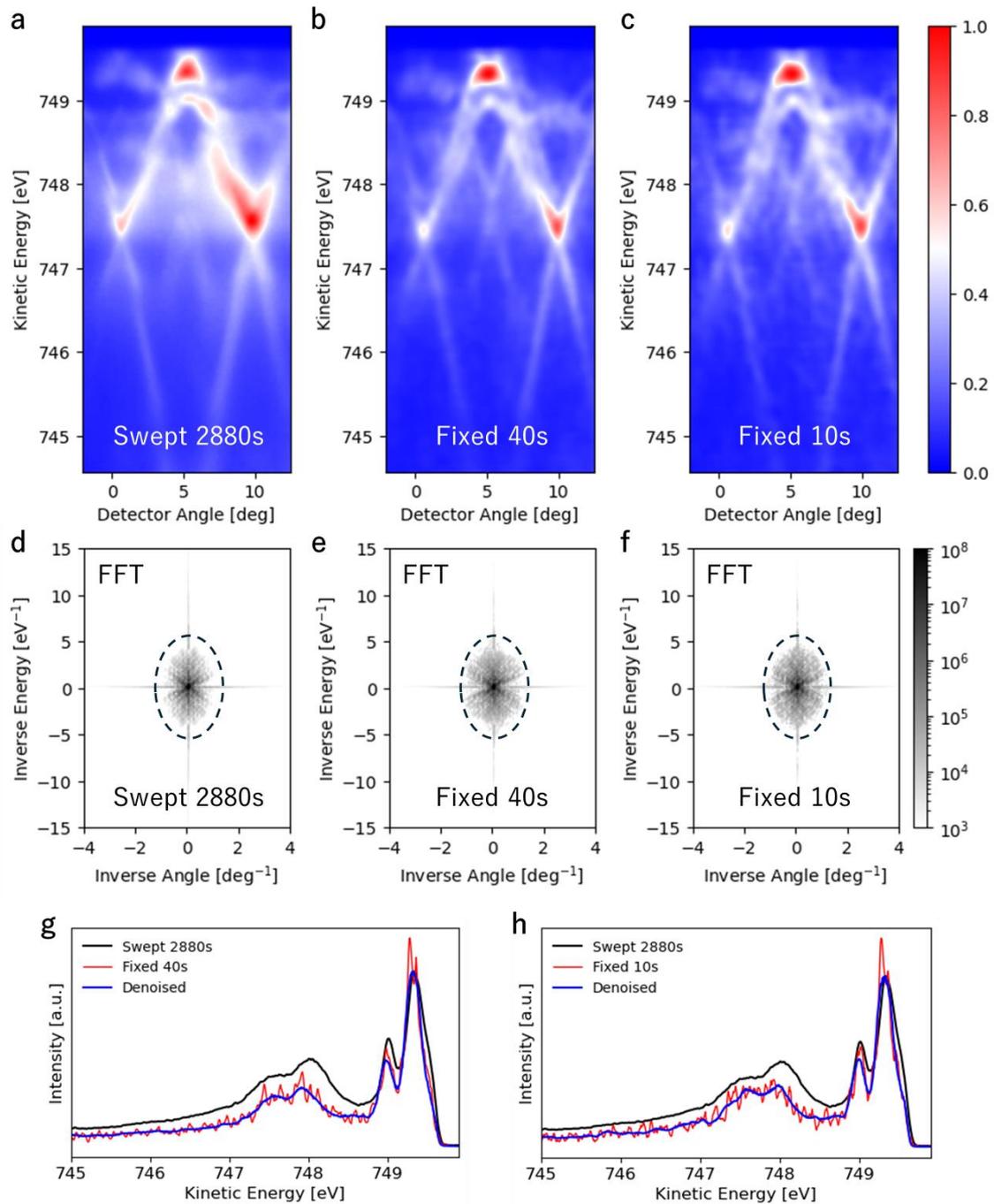

**Fig. 3: Comparison with the conventional swept mode. a**, Swept mode ARPES image acquired over 2880 s. **b**, **c**, Denoised results from fixed mode ARPES images acquired over 40 s and 10 s, respectively. **d**–**f**, Corresponding FFT images. **g**, **h**, Comparison of EDCs at 5.2°.

**Comparison with the conventional swept mode**

To validate our proposed method, we compared its performance with the conventional swept mode. For this benchmark, a swept mode ARPES image of the same sample was acquired with a typical measurement time of 2880 s (Fig. 3a). The intensity of each ARPES image was normalized to a range of 0 to 1. In the swept mode ARPES image, grid-like features are absent because the electric field sweep averages out such artifacts along the energy axis. However, the X-shaped band structure near the center (~ 5°, 747.5 eV) exhibits weak contrast and appears blurry. In contrast, the denoised result from the 40 s fixed mode measurement (Fig. 3b) reveals the X-shaped band structure with substantially greater clarity. Other band structures also show good qualitative agreement with the swept mode result.

Figure 3c presents the denoised result from an even shorter 10 s fixed mode measurement. Although some residual noise remains due to the extremely low S/N of the original image, the primary band structures observed in the swept mode are successfully reproduced. Notably, the central X-shaped feature appears sharper than that in the 2880 s swept mode image, demonstrating that even a 10 s measurement can provide a qualitative understanding of the band structures.

The corresponding FFT images (Figs. 3d–f) show no spots associated with grid structures, confirming the absence of the grid artifacts in the swept mode and denoised fixed mode results. While the FFT images appear broadly similar, the denoised fixed mode results exhibit additional high-frequency components, highlighted by the dashed ellipses. These components indicate finer band structures separated from the grid artifacts through deep prior-based denoising.

To investigate the blurriness of the X-shaped band structure in the swept mode, we compared the EDCs at 5.2° (Figs. 3g,h). Despite normalization by maximum intensity, the background (BG) in the swept mode is higher than that in the fixed mode across nearly the entire kinetic energy range. Around 747.5 eV, corresponding to the X-shaped band structure, this difference becomes particularly pronounced, indicating that the elevated BG in the swept mode measurement obscures the band structure. To verify that this BG difference is intrinsic rather than an artifact of deep prior-based denoising, we also included the EDC from the original fixed mode ARPES image for reference. A plausible explanation for the higher BG in the swept mode is the addition of the energy-averaged grid artifacts to the BG.

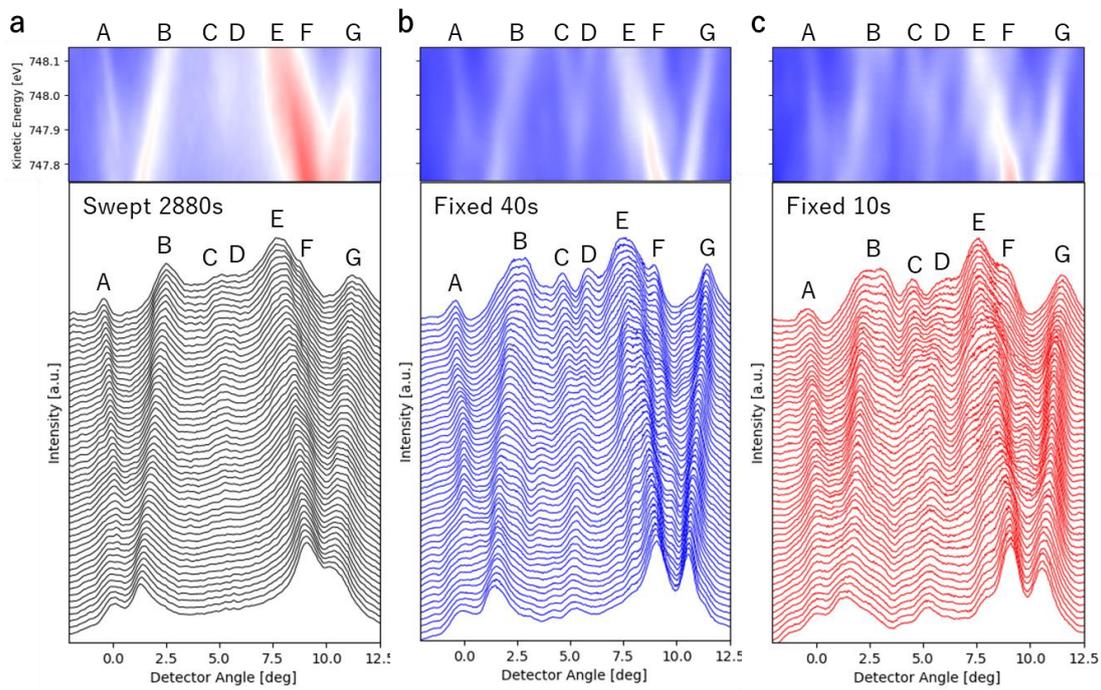

**Fig. 4: Comparison of MDCs. a**, Swept mode MDCs acquired over 2880 s. **b**, **c**, Denoised fixed mode MDCs acquired over 40 s and 10 s, respectively. The kinetic energy range spans 747.75–748.14 eV. Corresponding ARPES images are also shown.

Figure 4 presents a comparison of MDCs for a detailed investigation of the band structures. The kinetic energy range extends from 747.75 eV, near the center of the X-shaped band structure, to 748.14 eV, where the feature begins to split into two bands. In the swept mode, band structures labeled A through G are identified (Fig. 4a). However, features C and D, which were unclear in the ARPES image, are almost merged with the BG, making the two bands difficult to resolve. In contrast, the result from the 40 s fixed mode measurement clearly resolves all features A through G, with peaks C and D distinctly separated (Fig. 4b). In the 10 s fixed mode result, features A through G remain observable. However, the MDCs are somewhat distorted by noise (Fig. 4c). These results indicate that, for quantitative analysis of band structures using MDCs, a 40 s fixed mode measurement is preferable to a 10 s acquisition.

## Discussion

We demonstrate that deep prior-based denoising drastically reduces ARPES measurement time compared with conventional swept mode measurements. Beyond accelerating data acquisition, this approach produces ARPES images with higher clarity than those obtained in the swept mode. For qualitative determination of band structures from ARPES images, the measurement time can be reduced from 2880 s to just 10 s, representing nearly a 300-fold improvement in efficiency. For quantitative analysis using EDCs and MDCs, the time can be reduced from 2880 s to 40 s, achieving approximately a 70-fold enhancement.

Conventional swept mode measurements attempt to mitigate the grid artifacts experimentally, but these effects are never eliminated; the energy-averaged grid signal remains in the BG. Since the averaging does not occur along the angular direction, the angular dependence of the grid signal continues to influence the ARPES image, complicating the observation of the intrinsic band structure. In contrast, fixed mode measurements provide ARPES images with higher statistical precision and lower BG.

By combining deep prior-based denoising with the fixed mode acquisition, ultra-efficient ARPES measurements surpass conventional swept measurements in speed and resolution. This approach enables high-speed, high-resolution mapping of the 3D band structures in momentum space, bringing about a breakthrough for materials science. Leveraging differences in learning speeds between signals and noise, our deep prior-based denoising framework is broadly applicable, not only to ARPES but also to other advanced scientific imaging and metrology techniques, including computed tomography, X-ray scattering, and neutron scattering. Overall, our study establishes a benchmark for applying deep learning to scientific measurements in domains where data acquisition is inherently limited.

## Methods

### ARPES experiment

ARPES data for $CeRu_2Si_2$ were acquired at SPring-8 BL25SU using a µ-focused soft X-ray ARPES system equipped with a DA30 electrostatic hemispherical analyzer (Scienta Omicron)[17,18]. $CeRu_2Si_2$ was selected as an ideal sample for demonstrating deep prior-based denoising because its well-defined band structure has been previously characterized at the same beamline[19]. All measurements were performed at a temperature of 77 K. The incident photon energy was set to 750 eV, corresponding to the Γ-X line in the three-dimensional Brillouin zone, with an overall energy resolution of 90 meV.

### Deep prior-based denoising

For an inverse problem, such as obtaining a noise-free image $x^*$ from an experimentally observed image $x_{obs}$, a regularization-based approach is generally employed:

$$x^* = \min_{x} E(x, x_{obs}) + R(x), \quad (2)$$

where, $E(x, x_{obs})$ is the data fidelity term, representing the similarity between the images, commonly quantified using the MSE. The regularization term $R(x)$ encodes the intrinsic properties of the image, such as TV regularization, which enforces similarity between adjacent pixels. Prior studies have shown that such explicit regularization can be replaced by an implicit prior embedded within the neural network architecture itself[13].

Considering a neural network with parameters $\boldsymbol{\theta}$ that generates an output image $\boldsymbol{x} \in R^{1 \times H \times W}$ from a random input image $\boldsymbol{z} \in R^{1 \times H \times W}$, the relationship can be expressed as $\boldsymbol{x} = f_{\boldsymbol{\theta}}(\boldsymbol{z})$. Equation (2) can be formulated without the explicit regularizer as

$$\boldsymbol{x}^* = f_{\boldsymbol{\theta}^*}(\boldsymbol{z}). \qquad (3)$$

Where the optimal parameters $\boldsymbol{\theta}^*$ are obtained by optimizing $\boldsymbol{\theta}$ with respect to the observed image $\boldsymbol{x}_{obs} \in R^{1 \times H \times W}$ as

$$\boldsymbol{\theta}^* = \arg\min_{\boldsymbol{\theta}} E\{f_{\boldsymbol{\theta}}(\boldsymbol{z}), \boldsymbol{x}_{obs}\}. \qquad (4)$$

In this study, the observed image $\boldsymbol{x}_{obs}$ is modeled as the sum of the intrinsic ARPES signal, the grid artifacts, and random statistical noise. Our objective is to extract only the intrinsic ARPES signal. However, directly optimizing Eq. (4) would reconstruct the observed image $\boldsymbol{x}_{obs}$ perfectly, making it impossible to separate the grid artifacts and noise. We address this challenge by implementing early stopping. Prior work has shown that neural networks initially ignore noise during learning[13]; therefore, halting the learning process before the network begins fitting noise enables effective separation.

To separate the signal from the grid artifacts, a clear guideline is not available. We therefore exploited the spectral bias of CNNs[16], wherein learning progresses from low-frequency to high-frequency components. Assuming a weak correlation between the intrinsic signal and the grid artifacts, we designed a deep learning model (Table 1) that sequentially learns features, beginning with the ARPES signal. In our model (Fig. 2), the network primarily learns the ARPES signal during the first 400 iterations while ignoring both the grid artifacts and noise. This learning process allows the extraction of the pure ARPES signal within the plateau region, where learning temporarily stagnates. The optimal iteration number for early stopping can be determined from the learning curve, the output images, and their corresponding FFTs, as illustrated in Fig. 2.

**TABLE 1. Neural network architecture.**

| Encoder network |
| --- |
| Input z |
| Conv2d, BN, 5 × 5 × 16, stride = 1, lReLU |
| Max Pooling, Conv2d, BN, 5 × 5 × 32, stride = 1, lReLU |
| Max Pooling, Conv2d, BN, 5 × 5 × 64, stride = 1, lReLU |
| Max Pooling, Conv2d, BN, 5 × 5 × 128, stride = 1, lReLU |
| Decoder network |
| ConvTranspose2d, Conv2d, BN, 5 × 5 × 64, stride = 1, lReLU |
| ConvTranspose2d, Conv2d, BN, 5 × 5 × 32, stride = 1, lReLU |
| ConvTranspose2d, Conv2d, BN, 5 × 5 × 16, stride = 1, lReLU |
| Conv2d, BN, 1 × 1 × 1, stride = 1, sigmoid |

**Supplementary Information**

**Evaluation of the effect of the random seed**

All demonstrations of deep prior-based denoising in the main text were performed using a random seed of 123. To evaluate the effect of seed variability, we assessed the performance of the method under different random seeds. Supplementary Figure 1 shows an overlay of the MSE learning curves for 200 seeds (1–200). The plateau region appears as a dense blue band, indicating that a plateau consistently emerges between approximately 200 and 500 iterations across most random seeds. This behavior aligns well with the results in Fig. 2a of the main text, confirming that deep prior-based denoising operates robustly and reproducibly regardless of the initial random seed.

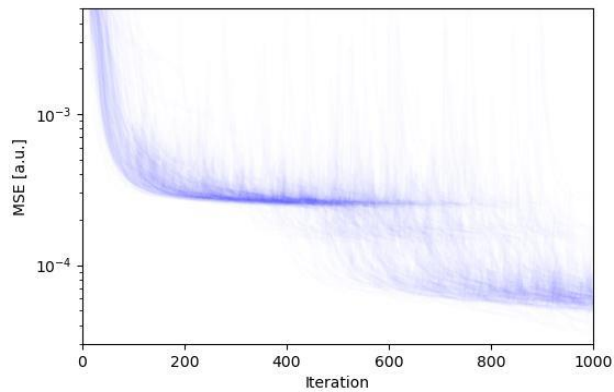

**Supplementary Figure 1: Effect of the random seed on performance.** Overlay of MSE learning curves generated from 200 different random seeds.